\documentclass[10pt]{article}

\usepackage{amsmath}
\usepackage{amssymb}

\usepackage{graphicx}

\usepackage{cite}

\usepackage{color}

\usepackage[labelfont=bf,labelsep=period,justification=raggedright]{caption}

\makeatletter
\renewcommand{\@biblabel}[1]{\quad#1.}
\makeatother

\date{}

\begin{document}

\begin{flushleft}
{\Large
\textbf{Heterogeneity Involved Network-based Algorithm Leads to Accurate and Personalized Recommendations}
}
\\
Tian Qiu$^{1,\ast}$,
Tian-Tian Wang$^{1}$,
Zi-Ke Zhang$^{2,3,4}$,
Li-Xin Zhong$^{5}$,
Guang Chen$^{1}$
\\
\bf{1} School of Information Engineering, Nanchang Hangkong
University, Nanchang, 330063, P. R. China
\\
\bf{2} Institute of Information Economy, Hangzhou Normal University, Hangzhou, 310036, P. R. China
\\
\bf{3} Web Sciences Center, University of Electronic Science and Technology of China, Chengdu, 610054, P. R. China
\\
\bf{4} Beijing Computational Science Research Center, Beijing, 100084, P. R. China
\\
\bf{5} School of Finance, Zhejiang University of Finance and Economics, Hangzhou, 310018, P. R. China

$\ast$ E-mail: tianqiu.edu@gmail.com

\end{flushleft}

\section*{Abstract}

Heterogeneity of both the source and target objects is taken into account in a network-based algorithm for the directional resource transformation between objects. Based on a biased heat conduction recommendation method (BHC) which considers the heterogeneity of the target object, we propose a heterogeneous heat conduction algorithm (HHC), by further taking the source object degree as the weight of diffusion. Tested on three real datasets, the \emph{Netflix}, \emph{RYM} and \emph{MovieLens}, the HHC algorithm is found to present a better recommendation in both the accuracy and personalization than two excellent algorithms, i.e., the original BHC and a hybrid algorithm of heat conduction and mass diffusion (HHM), while not requiring any other accessorial information or parameter. Moreover, the HHC even elevates the recommendation accuracy on cold objects, referring to the so-called cold start problem, for effectively relieving the recommendation bias on objects with different level of popularity.

\section*{Introduction}
The development of internet has made it easy to access information, which also brought about great convenience for our daily life. On the other hand, when facing various information, one is also puzzled how to get what he/she really wants. As a powerful tool, recommender system emerges to help people out of the overloaded information, which therefore attracts great interest of scientists from different disciplines \cite{ado05}, including physicists\cite{lv12,zha11}.

Different algorithms have been proposed, and achieved considerable progress. One of the most widely applied algorithm is the so-called collaborative filtering algorithm\cite{gol92,sch07}, which can also be divided into the memory-based \cite{breese98, nakamura99, delgado99} and model-based collaborative fitering\cite{getoor99, hofmann99, marlin03}. Another Line is the content-based algorithm \cite{paz07}. Various extensive algorithms have been comprehensively investigated \cite{bal97,gol01,hof04,ble03,lau06}, such as the hybrid algorithms of collaborative filtering and content-based algorithm.

A recommender system can be taken as a complex system composed of some interactive units \cite{zho08,zha10a,sha10,liu09b,liu12a,liu11b,zen12,zho07,zho10}. The complex interactions may affect the user's activity. For example, it makes a movie well known by advertising for it via medias like internet or television. After watching the movie, audiences may feedback favorable comments or negative criticism, which on the other hand should have an impact on the potential audiences, as well as the popularity of the movie. The comments made by users to an extent reflect the individual preference. Now, it has accumulated a large amount of historical data of the users' past activities, which makes it possible to design effective recommendation algorithms, and provide personalized recommendation for users by analyzing the data.

Physicists devote to the scientific program of pushing the personalized  recommendation by applying the concepts from statistical physics. Different physical-concept-based recommendation algorithms have been proposed \cite{zho07,zha07,zho09,ren08,zho10,zho11,qiu13a,lv11}, and have presented great advantages in personalized recommendation. A typical example is the standard heat conduction algorithm (HC) \cite{zha07}, which generates a highly personalized but less accurate recommendation. To improve the recommendation accuracy, many improved algorithms have been studied from different perspectives \cite{zho10,qiu11a,lv11}, e.g., introducing accessorial information such as social tags\cite{zha10a,kim10tag,zha12a}. Among those improved algorithms, a biased heat conduction algorithm (BHC)\cite{liu11} is reported as an excellent one, for prominently enhancing both the recommendation accuracy and diversity, well resolving the long-standing challenge of accuracy-diversity dilemma, without any aid of accessorial information.

Most heat-conduction-incorporated algorithms describe the recommender system as a user-object bipartite network, where objects affect each other via the linked users. Similarly as the directional characteristic of the heat flow, the influence between objects are also found to be directional. If a resource is assigned to each object, the object disseminating resource is named as source object, while the object receiving resource is called target object. The standard heat conduction method does not consider the object heterogeneity. However, most network-based recommender systems show heterogeneous structures. Without effectively eliminating the recommendation bias on objects with different level of popularity induced by the system heterogeneity, algorithms generally cannot achieve an excellent performance in both accuracy and diversity. To alleviate the recommendation bias, the biased heat conduction algorithm is proposed\cite{liu11}, which takes the object heterogeneity into account, but unilaterally for the target objects, whereas overlooks the heterogeneity of the source objects. In fact, the source object also presents heterogeneous characteristic. For example, some objects are popular with large degrees, while others are cold with small degrees. Obviously, the popular objects would contribute to much more users than the cold objects. Therefore, it is essential to take the heterogeneity of both the source and target objects into consideration in the algorithm design.

In this article, based on the biased heat conduction algorithm\cite{liu11}, we further introduce the source object degree as the diffusion weight, and propose the heterogeneous heat conduction algorithm (HHC). Tested on three real datasets, the \emph{Netfilx}, \emph{RYM}, and \emph{MovieLens}, the HHC is found to present an excellent performance in both the recommendation accuracy and diversity, even more advantageous than the original biased heat conduction algorithm, and also an excellent hybrid algorithm of heat conduction and mass diffusion (HHM)\cite{zho10}, for effectively relieving the recommendation bias induced by the system heterogeneity. Moreover, the HHC meanwhile elevates the recommendation accuracy on cold objects, better resolving the so-called cold start problem.

\section*{Materials and Methods}

A recommender system can be characterized by a bipartite graph composed of the user set $U$ containing $m$ users, and the object set $O$ containing $n$ objects. If an object $o_\alpha$ is collected by a user $u_i$, then add a link between them. The relation between the user and the object can be described by an adjacent
matrix $A=\{a_{i\alpha}\}$, with $a_{i\alpha}$ to be 1 if there is a link between the user-object pair, otherwise, to be 0.

The standard heat conduction method is firstly proposed by Zhang et al \cite{zha07}, by introducing the heat conduction analogous process into the recommender systems. Assume each object has an initial resource. The resource would flow between different objects directionally like the heat. That is to say, the resource would spread from the source object to target object through the linked user, and all the objects would achieve a final resource. The resource diffusion process can be described by,

\begin{equation}
\textbf{f'}=\textbf{W}\textbf{f}
\end{equation}

where $W$ is the transformation matrix, characterizing the resource diffusion process from the source object to target object. $\textbf{f}$ is the initial resource of the object, and the $\textbf{f'}$ is the final resource. For each user, rank his/her uncollected objects in the decreasing order of the final resource, and then recommend the top $L$ objects to the user. The simplest way to assign the initial resource to object is to set 1 or 0, on the basis of whether the object is collected by the user or not. The initial resource of user $u_{i}$ to object $o_{\alpha}$ can be described as,

\begin{equation}
\vec{f}_{0(\alpha)}^{i}=a_{i\alpha}
\end{equation}

If an object is collected by the user $u_i$, its initial resource is assigned to be 1, otherwise, to be 0. All the following algorithms are based on the simplest assigning way of initial resource.

The process to redistribute the resources, here represented by the transformation matrix $W$, therefore plays a key role in the recommendation algorithm. An illustration of the diffusion process of the standard heat conduction algorithm (HC) is shown in Fig. 1 (a). At first, the particular user $i$ indicated by the solid circle receives an average level resource from his/her neighboring objects. Here the user $i$ has two neighbors, the first and the fourth neighbors, therefore can get the average resource of $1$. Then the objects again get the average resources from all their neighboring users. The transformation matrix of the HC method can be formulated by,

\begin{equation}
W_{\alpha\beta}^{HC}=\frac{1}{k_{\alpha}}\sum
_{j\in U}\frac{a_{\alpha j}a_{\beta j}}{k_{j}}, \label{eq.3}
\end{equation}

where $k_{\alpha}$ is the degree of object $o_{\alpha}$, and $k_{j}$ is the degree of user $u_{j}$. The transformation matrix $W$ is found to be asymmetrical, i.e., the influence of the object $o_{\beta}$ to $o_{\alpha}$ is different from that of the object $o_{\alpha}$ to $o_{\beta}$. For the directional influence from the object $o_{\beta}$ to $o_{\alpha}$ in $W_{\alpha\beta}$, we notate the object $o_{\beta}$ as the source object, and $o_{\alpha}$ as the target object, respectively. The HC method assigns more priority to the low-degree objects, which leads to a highly personalized, but less accurate recommendation.

Generally, enhancing recommendation accuracy inhibits recommendation diversity. Effectively solving the accuracy-diversity dilemma has been a long-standing challenge of recommender systems. Recently, Liu et al proposed a biased heat conduction (BHC) method \cite{liu11}, by taking the heterogeneity of the target object into account, with its transformation matrix formulated by,

\begin{equation}
W_{\alpha\beta}^{BHC}={\frac{1}{k_{\alpha}^{\gamma}}}\sum_{j\in U}\frac{a_{\alpha j}a_{\beta j}}{k_{j}}, \label{eq.4}
\end{equation}

where $\gamma$ is a tunable parameter. Compared with a number of network-based recommendation algorithms, the BHC shows a great advantage in both the recommendation accuracy and diversity \cite{liu11}. In our study, based on the BHC, we further consider the heterogeneous effect of the source objects, and propose the heterogeneous heat conduction method (HHC).

In the network-based recommender systems, due to the heterogeneity of the source objects, the contribution of the source objects to users should be quite different. For instance, the popular objects usually have a big degree for widely collected by users, while the cold objects generally own a small degree. Hence, the popular objects should contribute to much more users than the cold objects. Using the source object degree as the diffusion weight, the transformation matrix of the HHC can be formulated by,

\begin{equation}
W_{\alpha\beta}^{HHC}={\frac{1}{k_{\alpha}^{\gamma}}}\sum_{j\in U}\frac{a_{\alpha j}a_{\beta j}}{k_{j}k_{\beta}}, \label{eq.4}
\end{equation}

In order to show the advantages of the HHC method, we compare it with another excellent method, i.e., the hybrid method of heat conduction and mass diffusion (HHM) \cite{zho10}, which is outstanding in both the recommendation accuracy and diversity. So far, the network-based recommendation algorithm that outperforms the HHM is still rarely reported. Although there are few algorithms are reported to outperform the HHM in some aspect, they usually introduce additional accessorial information or parameter \cite{zha10b,liu12}. Therefore, comparing the HHC with the HHM can give a solid evidence of evaluating the algorithm performance, since the HHC does not require any accessorial information or extra parameter.

We firstly introduce the so-called mass diffusion method (MD), with an example of the diffusion process shown in Fig. 1 (b). At first, each object distributes the resource to its neighboring users with an equal probability. Then the user again redistributes all his/her resource to his/her neighboring objects, also with the equal probability. By summing up all the resources from their neighboring users, the objects then obtain their final resources. The transformation matrix of the MD algorithm is formulated by,

\begin{equation}
W_{\alpha\beta}^{MD}=\frac{1}{k_{\beta}}\sum
_{j \in U}\frac{a_{\alpha j}a_{\beta j}}{k_{j}}, \label{eq.2}
\end{equation}

The MD method assigns more priority to the popular objects, since objects receive resources from all their neighboring users in the last step of diffusion. Such a diffusion pattern results in an excellent recommendation accuracy, yet a relatively poor diversity.

The hybrid method of heat conduction and mass diffusion (HHM) takes an advantage of the two processes \cite{zho10}, with its transformation matrix formulated by,

\begin{equation}
W_{\alpha\beta}^{HHM}=\frac{1}{k_{\alpha}^{1-\gamma}k_{\beta}^{\gamma}}\sum
_{j \in U}\frac{a_{\alpha j}a_{\beta j}}{k_{j}}, \label{eq.4}
\end{equation}

where $\gamma \in [0,1]$. When tuning the parameter $\gamma$ to a proper value, the HHM method presents a high recommendation efficiency in both the accuracy and diversity.

\section*{Metrics}
The most important aspect of assessing the performance of a recommendation algorithm is the recommendation accuracy. To give a solid evaluation of the accuracy, we use three indicators to present the accuracy performance, i.e., the ranking score $\langle RS \rangle$, precision $P$, and recall $R$. On the other hand, to show how personalized the algorithm, we use the novelty and diversity to evaluate the personalization performance.


1. \textbf{Ranking Score ($\langle RS \rangle$)}\cite{zho07}\textbf{.-}In the network-based recommender system, if an object is collected by a user, i.e., there is a link between the user and the object, we take the object preferred by the user. The ranking score $RS_{\alpha i}$ then quantifies how the deleted link of object $o_{\alpha}$ to user $u_{i}$ in the test set rank in all $u_{i}$'s deleted links, which is defined as,

\begin{equation}
RS_{\alpha i}=\frac{p_{\alpha}}{n-k_{i}}.
\end{equation}

where $n$ is the number of all objects, $k_{i}$ is the degree of the user $u_{i}$, and $p_{\alpha}$ is the position of the recommended object $o_{\alpha}$ located in all the uncollected objects of the user $u_{i}$. Obviously, the smaller the $RS_{\alpha i}$, the higher rank of the deleted link, and the more accurate the algorithm. The average ranking score $\langle RS \rangle$ is taken an average of $RS_{\alpha i}$ over all the deleted links.

2. \textbf{Precision ($P$)}\cite{her04}\textbf{.-}The recommendation precision $P$ evaluates how the deleted links are recovered, which is defined as

\begin{equation}
P=\frac{1}{m}\frac{\sum_{i=1}^{m}q_{iL}}{L}, \label{eq.7}
\end{equation}
where $q_{iL}$ is the number of the user $u_{i}$'s deleted links contained in the top $L$ recommended object list. The higher the precision, the more accurate the recommendation, and vice versa.

3. \textbf{Recall ($R$)}\cite{her04}\textbf{.-} The recall $R$ is defined as

\begin{equation}
R=\frac{1}{m}\sum_{i=1}^{m}\frac{q_{iL}}{l_{i}}, \label{eq.6}
\end{equation}
where $q_{iL}$ is the number of the user $u_{i}$'s deleted links contained in the top $L$ recommended object list, $l_{i}$ is the
number of the user $u_{i}$'s deleted links in the test set. The higher the recall, the more accurate the recommendation, and vice versa.


4. \textbf{Novelty ($NL$).-} The novelty indicates how unexpected the recommended objects to the user. Here we use the average degree of the objects in the recommendation list to quantify the novelty, which is defined as,

\begin{equation}
NL=\frac{1}{mL}\sum_{i=1}^{m}\sum_{o^{i}_{\alpha} \in O^{i}_{R}}^{}k_{o^{i}_{\alpha}}, \label{eq.9}
\end{equation}

where $O^{i}_{R}$ is the object set of the user $u_{i}$'s recommendation list. The smaller the $NL$, the more novel the recommendation to the user, and vice versa.

5. \textbf{Hamming Distance ($H$)).-} The recommendation diversity is quantified by the Hamming distance $H$ of the recommendation lists of two different users, which is defined as,

\begin{equation}
H=\frac{2}{m(m-1)}\sum_{i=1}^{m}\sum_{j=i+1}^{m}(1-\frac{\emptyset_{i} \bigcap \emptyset_{j}}{L}), \label{eq.8}
\end{equation}
where $\emptyset_{i} \bigcap \emptyset_{j}$ is the number of the common objects recommended for the user $u_{i}$ and $u_{j}$ in the top $L$
recommendation list. The Hamming distance $H$ therefore evaluates how different the recommendation lists of users. The higher the $H$, the less in common the recommended objects for different users, i.e., the more diverse the algorithm, and vice versa.

\subsection*{Data}
In order to provide a comprehensive understanding of the algorithms, we employ the algorithms on three empirical datasets, i.e., the \emph{Netfilx}, \emph{RYM}, and \emph{MovieLens}. The \emph{Netflix} and \emph{MovieLens} are both five-level rating movie systems, and the \emph{RYM} is a ten-level rating music system. The \emph{Netflix} is randomly selected from the huge dataset of the \emph{Netflix} Prize, the \emph{MovieLens} is downloaded from the web site of GroupLens Research (http://grouplens.org), and the \emph{RYM} is downloaded from the music rating web site RateYourMusic.com.

To construct the bipartite network, a link between a user and an object is added if the rating of the user to the object is no less than three for the \emph{Netflix} and \emph{MovieLens}, and no less than six for \emph{RYM}. The basic statistics of the three datasets is summarized in table 1. The sparsity of the dataset is defined as the number of links proportional to the total number of the user-object links. To test the algorithm performance, all the links of the network are split into two subsets, i.e., the training set and the test set. Randomly delete $10\%$ links as the test set to test the performance of the algorithm, and remain the rest $90\%$ links as the training set to make predictions for users.

\section*{Results and Discussion}

To evaluate the algorithm performance, we compare the results of the HHC with the excellent BHC and HHM algorithms, with the results being the average over six runs. The recommendation accuracy results are presented by the indicators of ranking score $\langle RS \rangle$, precision $P$, and recall $R$. For all the three algorithms, one tunable parameter is introduced. To obtain the optimal value of the tunable parameter $\gamma$, we investigate the ranking score on $\gamma$. As shown in Fig. 2, for the HHC, the minimal value of the ranking score is obtained at $\gamma = 0.68$, 0.61 and 0.74 for the \emph{Netflix}, \emph{RYM} and \emph{MovieLens}, respectively. Similar procedure is also carried for the BHC and HHM method. For the BHC method, the minimal value of the ranking score is obtained at $\gamma = 0.85$, 0.80 and 0.85 for the \emph{Netflix}, \emph{RYM} and \emph{MovieLens}, and for the HHM method, the minimal value of the ranking score is obtained at $\gamma = 0.17$, 0.25 and 0.17 for the \emph{Netflix}, \emph{RYM} and \emph{MovieLens}. The following results are obtained at the optimal value of $\gamma$.

\subsection{Accuracy and Personalization of Recommendation}

The performance of the HHM, BHC and HHC algorithms are summarized in table 2. For the ranking score $\langle RS \rangle$ and the precision $P$, the HHC algorithm outperforms both the HHM and BHC for all the three datasets. The recall $R$ of the \emph{RYM} and \emph{MovieLens} of the HHC is also more advantageous than the HHM and BHC. It indicates that the HHC is more effective in recommendation accuracy.

To quantitatively examine how much the HHC outperforms the HHM and BHC, we define an improvement percentage as $\delta_{ALG}= (Q_{SCL}-Q_{ALG})/Q_{ALG}$, where the subhead $ALG$ refers to the investigated algorithm, and the $Q_{ALG}$ is the value of the indicator. The results of the improvement percentage are summarized in table 3. Taking the \emph{Netflix} as an example, the improvement percentage of the ranking score of the HHC against the BHC and HHM is found to be as much as 14.6\% and 8.9\%. Such an improvement is very appreciable, since the HHC algorithm does not introduce any accessorial information or additional parameter.

The novelty $NL$ and diversity $H$ are also investigated to evaluate how personalized the algorithm. As shown in table 1, taking the \emph{Netflix} as an example, there is also a great improvement of the novelty for the HHC, with the improvement percentage of the HHC against the BHC and HHM as much as $8.2\%$ and $11.5\%$. Similarly, the diversity of the HHC method also shows a great advantage, with the improvement percentage of the HHC against the BHC and HHM to be $10.4\%$ and $12.5\%$. It suggests that the HHC is not only accurate, but also personalized.

To show how the personalized indicators evolve with the recommendation list length, we study the novelty $NL$ and diversity $H$ for different recommendation list lengths $L$. As shown in Fig. 3, for the \emph{Netflix} and \emph{RYM}, the novelty of the HHC is more advantageous than the HHM and BHC for a wide range of recommendation list lengths. For the \emph{MovieLens}, the novelty of the HHC is also more advantageous than the BHC, and is similar as the HHM algorithm. The diversity $H$ on the recommendation list lengths is shown in Fig. 4. It is observed, for all the three datasets, the diversity of the HHC is much more advantageous than that of both the HHM and BHC for all the investigated range of the recommendation list lengths. It further confirms the results in table 2 and 3.

Our results have shown that the HHC is more advantageous than the BHC and HHM in the recommendation accuracy, novelty, as well as diversity. To further test the robustness of the results, we adjust the proportion of the training set to the total data, and investigate the corresponding ranking score $\langle RS \rangle$, Hamming distance $H$ and novelty $NL$. As shown in Fig. 5, taking the \emph{Netflix} dataset as an example, it is observed that for all the investigated training set ratios, even for the sparse training data condition with its ratio as low as 0.4., the HHC outperforms the BHC and HHM, for all the three metrics, i.e., recommendation accuracy, novelty and diversity. It provides a further solid evidence of the great advantages of the proposed HHC method.

Why the HHC can achieve such an appreciable improvement in both aspects of accuracy and personalization? By analyzing the statistical property of the dataset, a great heterogeneity is observed for the data. As shown in Fig. 6 (a), a power-law-like probability distribution of the object degree is observed for the \emph{Netflix}, which indicates that a large proportion of the objects owns very small degrees, whereas a small fraction of the objects has big degrees. Assume that objects with big degrees are popular objects, and with small degrees are cold objects. Focusing on the recommendation of either popular objects or cold objects cannot achieve an effective recommendation simultaneously in the accuracy and diversity. Typical examples are the standard heat conduction method and the mass diffusion method. The standard heat conduction algorithm emphasizes the recommendation of cold objects, leading to a very personalized but less accurate recommendation, whereas the mass diffusion algorithm tends to recommend popular objects, generating an accurate yet relatively less personalized recommendation. Eliminating the recommendation bias on objects with different level of popularity induced by the system heterogeneity is essentially required for designing effective algorithms. By analyzing the degree distribution $p(k)$ for the objects in the top $L=50$ recommendation list shown in Fig. 6 (b), we find that the BHC and HHM methods have well resolved the recommendation for both the popular and cold objects. Compared with the BHC and HHM methods, the HHC obtains a better recommendation balance between the cold objects and popular objects, which to an extent explains why the HHC can further improve both the accuracy and personalization of recommendation.

To better understand the obtained results, we then analyze the eigenvalues of the transformation matrix $W$ for the BHC, HHM, and HHC. As shown in Fig. 7, the eigenvalues are displayed in the decreasing order of the values for the three methods. The large eigenvalues indicate that there exist advantageous components in the $W$. For the BHC method, several large eigenvalues are observed for the \emph{Netflix}, \emph{RYM} and \emph{MovieLens}. Similar large eigenvalues are also observed in the HHM method, but the values of the first several eigenvalues are smaller than those of the BHC. However, for the HHC method, nearly all the eigenvalues show very close values, i.e., all the components have a near effect. That is to say, regardless of the cold or popular objects, the HHC assigns an approximate weight for all the objects. The frequency distribution of the eigenvalues of the transformation matrix $W$ is shown in Fig. 8 for the \emph{Netflix}. The eigenvalue distribution of the HHC method indicates an apparently more homogeneous distribution than that of the BHC and HHM, with the largest eigenvalue of the HHC to be 0.35, far less than 2.68 of the BHC. It well accounts for the success of the HHC in relieving the recommendation bias. The results of the eigenvalue analysis of the transformation matrix $W$ are consistent with the observations from the degree distribution of the top $L$ recommended objects, which provides a deeper insight into understanding the underlying mechanism of the algorithm.

\subsection{Accuracy on Cold Objects}

From the analysis of the object degree distribution and the eigenvalues of the transformation matrix, the cold objects are observed to get more priority in the HHC, compared with the BHC and HHM. It reminds us of the so-called the 'cold-start' problem, referring to how to recommend the new objects or cold objects, i.e., the cold object start \cite{pak08}, or how to recommend objects to newly added users, i.e., the cold user start \cite{lam08}. Here we focus on the cold object start.

Due to lack of information, it is hard for users to be aware of the cold objects, resulting in the difficulty to recommend them effectively. However, the cold objects usually occupy a big proportion of the total objects. If we define the object with its degree no more than 10 as the cold object, there are as much as $49.59 \%$, $21.73\%$, and $41.26\%$ cold objects for the \emph{Netflix}, \emph{RYM} and \emph{Movielens}. Hence, how to solve the cold start problem is an important but challenging problem in recommender systems.

To present the recommendation accuracy on the cold objects, we investigate an object-dependent ranking score $\langle RS \rangle_{k}$, which is defined as the average ranking score over objects with the same value of degrees\cite{zho08}. As shown in Fig. 9, whereas showing similar recommendation accuracy on the popular objects with large degrees, the $\langle RS\rangle_{k}$ of the low-degree objects of the HHC is found to be smaller than that of both the BHC and HHM for all the three datasets. Focusing on objects whose degree are no more than 10, the $\langle RS \rangle _{k \leq 10}$ of the HHC is found to be much smaller than that of both the BHC and HHM, as shown in table 2. Taking the \emph{Netflix} as an example, the improvement percentage is as much as $23.0\%$ against the BHC, and $18.2\%$ against the HHM. Similar improvement is also found for the \emph{RYM} and \emph{MovieLens}.

Further, we investigate an object-dependent precision $P_{k}$ and an object-dependent precision $R_{k}$. $P_{k}$ is defined as $P_{k}=\frac{1}{m}\frac{\sum_{i=1}^{m}q_{iL}^{k}}{L}$, and $R_{k}$ is defined as $R_{k}=\frac{1}{m}\frac{\sum_{i=1}^{m}q_{iL}^{k}}{l_{i}^{k}}$,
where $q_{iL}^{k}$ is the number of the user $u_{i}$'s deleted links for objects with degree $k$ in the top $L$ recommended object list, and $l_{i}^{k}$ is the number of the user $u_{i}$'s deleted links for objects with degree $k$ in the test set. The larger the $P_{k}$ or $R_{k}$, the more accurate the recommendation on cold objects, and vice versa. As shown in table 2, both the $P _{k \leq 10}$ and $R _{k \leq 10}$ of the HHC greatly outperform those of the BHC and HHM. For the \emph{Netflix}, the improvement percentage of $P _{k \leq 10}$ is as much as $1019.4\%$ against the BHC, and $508.8\%$ against the HHM, and of $R _{k \leq 10}$ is $900.0\%$ against the BHC, and $511.1\%$ against the HHM. Similar improvement is also found for the \emph{RYM} and \emph{MovieLens}. It gives a strong evidence that the HHC better resolves the cold start problem than both the BHC and HHM.

By studying the three indicators of recommendation accuracy on cold objects, i.e., the $\langle RS \rangle_{k \leq 10}$, $P _{k \leq 10}$ and $R _{k \leq 10}$, for different ratios of the training set to the total data, the robustness of the results can be further confirmed. As shown in Fig. 10, taking the \emph{Netflix} as an example, the $R _{k \leq 10}$ of the HHC is much smaller than that of the BHC and HHM, and the $P _{k \leq 10}$ and $R _{k \leq 10}$ of the HHC are much higher than those of the BHC and HHM, for all the investigated ratios, even for the sparse training data with the ratio as low as 0.3. Similar behavior is also found for the \emph{RYM} and \emph{MovieLens}.

As mentioned above, users are seldom aware of the cold objects. So far, most studies try to solve the cold start problem with the aid of other accessorial information, such as the trust relationship \cite{jamali09trust}, and tags \cite{zha10a,zha10b,kim10tag,zha11}, or introducing additional parameter \cite{liu12}. However, these accessorial information or new parameter would make the recommender system more complex. Employing the source object degree as the weight of diffusion, and without introducing any accessorial information, the HHC further elevates the recommendation accuracy on cold objects, better resolving the cold start problem for fully taking the heterogeneity of objects into account.

\section*{Conclusion}
In conclusion, based on an excellent biased heat conduction method (BHC), we consider the heterogeneity of both the source and target objects, and propose the heterogeneous heat conduction recommendation algorithm (HHC), by taking the source object degree as the weight of diffusion. Without employing any other accessorial information or additional parameter, the HHC outperforms the original BHC method, and even a highly accurate and personalized hybrid method of heat conduction and mass diffusion (HHM), in both the recommendation accuracy and personalization. Moreover, the HHC further enhances the recommendation accuracy on cold objects, referring to the so-called cold start problem.

Due to the difficulty to alleviate the recommendation bias on objects with different level of popularity, improving recommendation accuracy usually inhibits recommendation diversity, causing the great challenge of accuracy-diversity dilemma. Especially, large amounts of new objects emerge in the online applications, which not only even intensifies the system heterogeneity, but also brings about the cold start problem. The HHC not only further improves both the recommendation accuracy and personalization, but also enhances the recommendation accuracy on cold objects, for greatly relieving the recommendation bias induced by the system heterogeneity. Our work might shed some new light on better understanding and designing effective recommendation algorithms.



\bibliography{article}

\begin{figure}[htb]
\centering
\includegraphics[width=8.5cm]{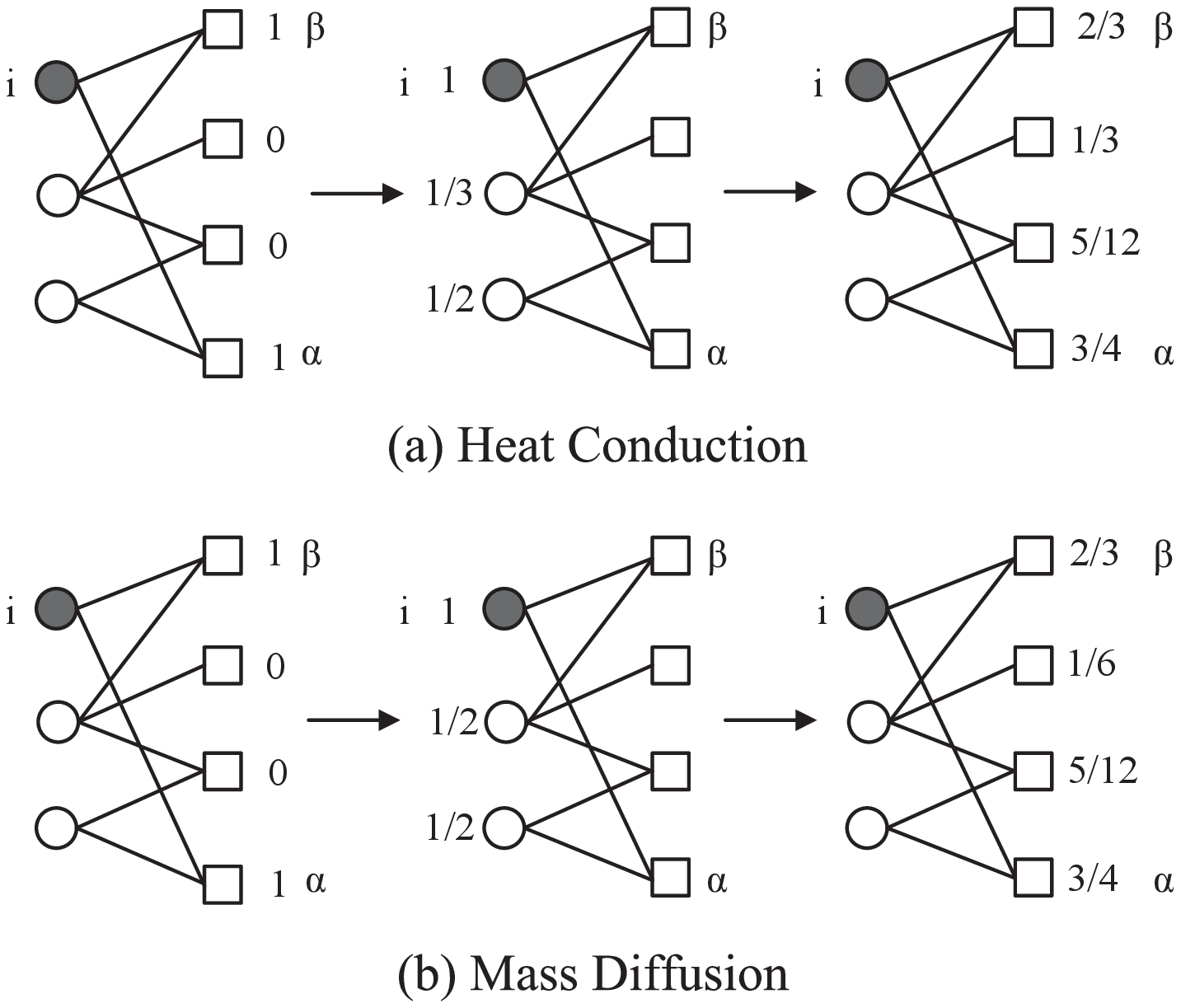}
\caption{\label{Fig:1} The illustration of the resource transformation. (a) for the heat conduction process, and (b) for mass diffusion process.}
\end{figure}

\begin{figure}[htb]
\centering
\includegraphics[width=8.5cm]{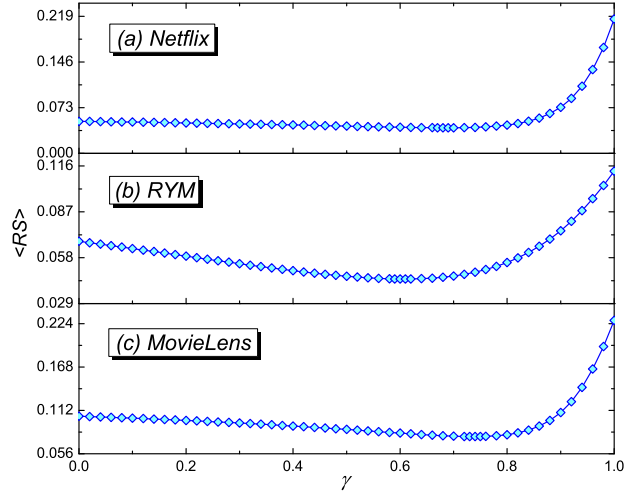}
\caption{\label{Fig:2} The ranking score $\langle RS \rangle$ on the tunable parameter $\gamma$ of the HHC algorithm. (a) for the \emph{Netflix}, (b) for the \emph{RYM} and (c) for the \emph{MovieLens}.}
\end{figure}

\begin{figure}[htb]
\centering
\includegraphics[width=8.5cm]{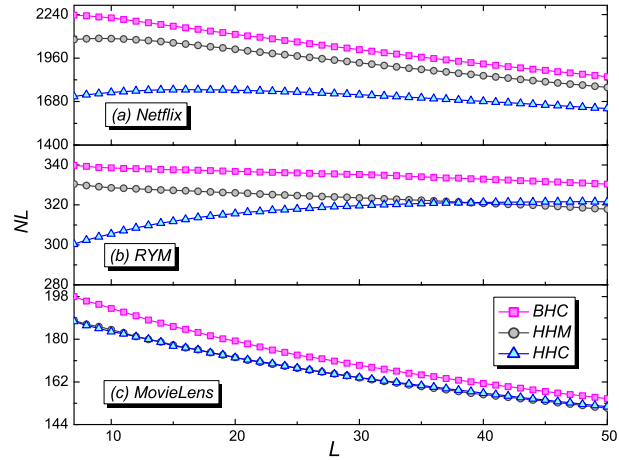}
\caption{\label{Fig:3} The novelty $NL$ on the recommendation list length $L$. (a) for the \emph{Netflix}, (b) for the \emph{RYM} and (c) for the \emph{MovieLens}. The magenta, grey and cyan lines are for the BHC, HHM and HHC methods, respectively.}
\end{figure}

\begin{figure}[htb]
\centering
\includegraphics[width=8.5cm]{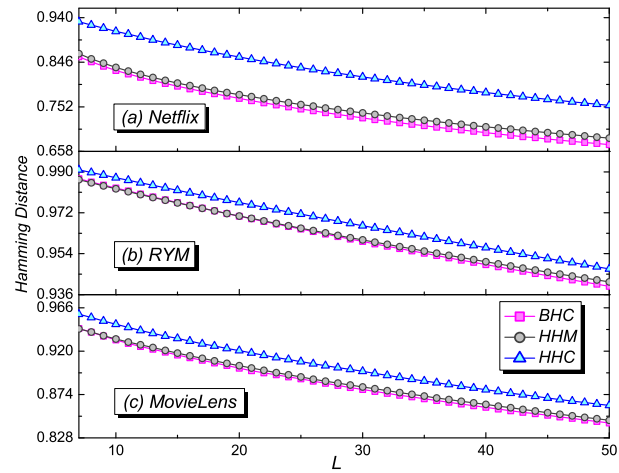}
\caption{\label{Fig:4} The Hamming distance $H$ on the recommendation list length $L$. (a) for the \emph{Netflix}, (b) for the \emph{RYM} and (c) for the \emph{MovieLens}. The magenta, grey and cyan lines are for the BHC, HHM and HHC methods, respectively.}
\end{figure}

\begin{figure}[htb]
\centering
\includegraphics[width=8.5cm]{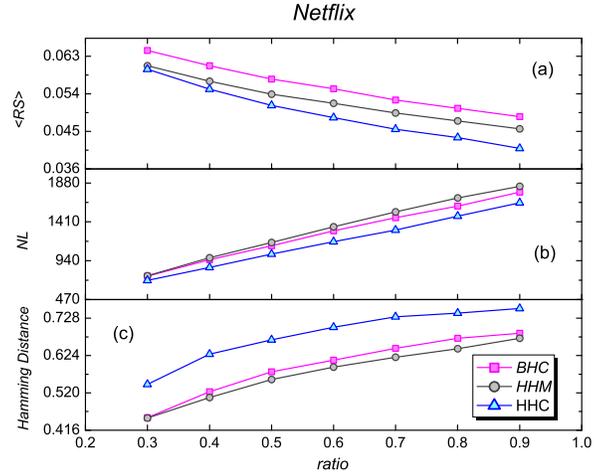}
\caption{\label{Fig:5} The ranking score $\langle RS \rangle$, novelty $NL$, and Hamming distance $H$ are displayed for different ratios of the training set to the total data in (a), (b) and (c) for the \emph{Netflix}, respectively. The magenta, grey and cyan lines are for the BHC, HHM and HHC methods, respectively.}
\end{figure}

\begin{figure}[htb]
\centering
\includegraphics[width=8.5cm]{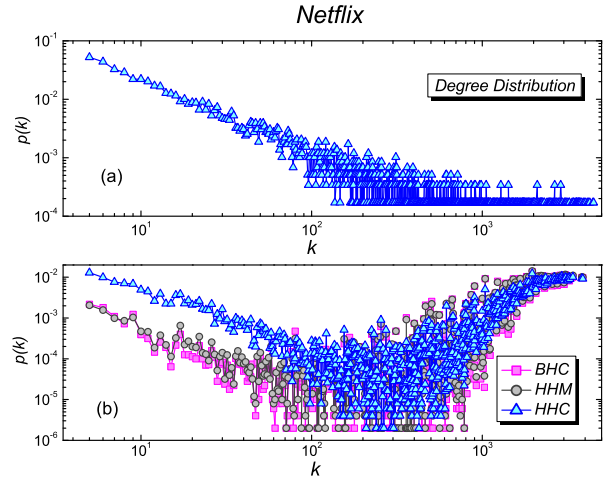}
\caption{\label{Fig:6} (a) The object degree distribution $p(k)$ of the \emph{Netflix}, and (b) the degree distribution $p(k)$ for the objects in the top $L =50$ recommendation list. In the subplot (b), the magenta, grey and cyan lines are for the BHC, HHM and HHC methods, respectively.}
\end{figure}

\begin{figure}[htb]
\centering
\includegraphics[width=8.5cm]{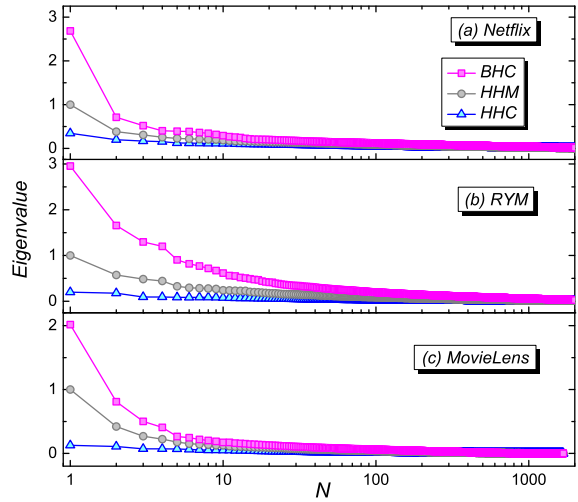}
\caption{\label{Fig:7} The eigenvalues of the transformation matrix $W$ are displayed in the decreasing order of the values. (a) for the \emph{Netflix}, (b) for the \emph{RYM} and (c) for the \emph{MovieLens}. The magenta, grey and cyan lines are for the BHC, HHM and HHC methods, respectively.}
\end{figure}

\begin{figure}[htb]
\centering
\includegraphics[width=8.5cm]{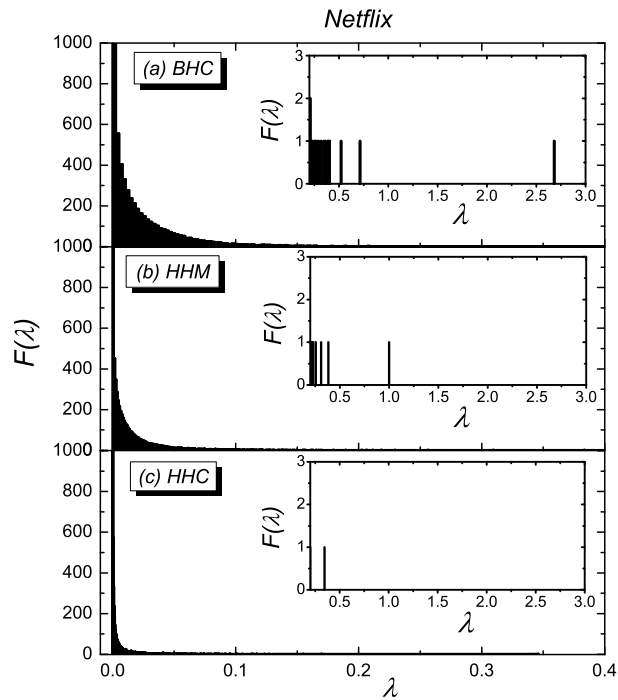}
\caption{\label{Fig:8} The frequency distribution of the eigenvalues $\lambda$ of the transformation matrix $W$ is displayed for the \emph{Netflix}. The inset shows the first several largest eigenvalues. (a) for the BHC method, (b) for the HHM method and (c) for HHC method.}
\end{figure}

\begin{figure}[htb]
\centering
\includegraphics[width=8.5cm]{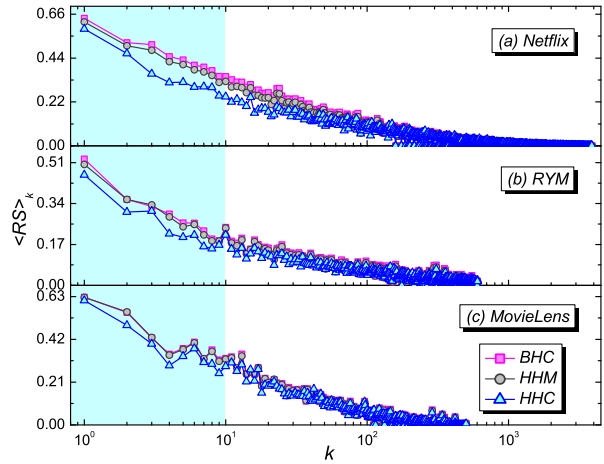}
\caption{\label{Fig:9} The object-dependent ranking score $\langle RS \rangle_{k}$ vs. the object degree $k$. (a) for the \emph{Netflix}, (b) for the \emph{RYM} and (c) for the \emph{MovieLens}. The magenta, grey and cyan lines are for the BHC, HHM and HHC methods, respectively.}
\end{figure}

\begin{figure}[htb]
\centering
\includegraphics[width=8.5cm]{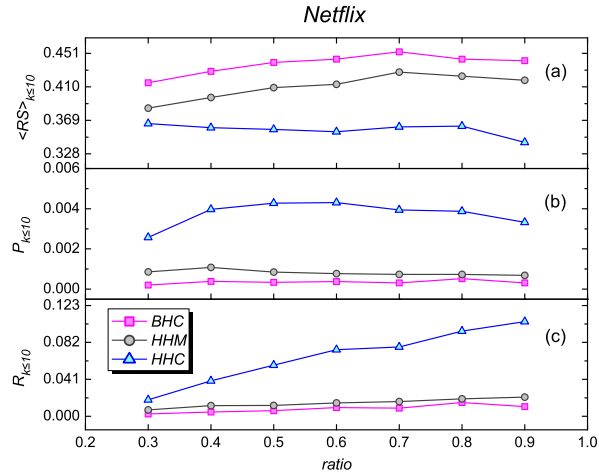}
\caption{\label{Fig:10} The object-dependent ranking score $\langle RS \rangle_{k\leq10}$, precision $P_{k\leq10}$, and recall $R_{k\leq10}$ are displayed for different ratios of the training set to the total data in (a), (b) and (c) for the \emph{Netflix}, respectively. The magenta, grey and cyan lines are for the BHC, HHM and HHC methods, respectively.}
\end{figure}

\section*{Tables}

\begin{table}[!ht]
\caption{
\bf{The statistics of the datasets.}}
\begin{center}
\begin{tabular}{|c|c|c|c|c|}
\hline  Dataset & Users & Objects & Links & Sparsity \\
\hline
\emph{Netflix}& 9999 & 5870 &  815917 & 1.39 \% \\
\emph{RYM} & 10159 & 5250 & 559634 & 1.05 \% \\
\emph{MovieLens} & 943 & 1682 &  100000 & 6.30 \% \\

\hline
\end{tabular}
\end{center}
\begin{flushleft}The statistics of the \emph{Netflix}, \emph{RYM}, and \emph{MovieLens} is displayed. From left to right, the
columns correspond to the name of the data, the number of users,
objects, and links, and the sparsity of the data.
\end{flushleft}
\label{tab:1}
\end{table}

\begin{table}[!ht]
\caption{
\bf{The performance of the HHM, BHC and HHC methods.}}
\begin{tabular}{|c|c|c|c|c|c|c|c|c|c|}
\hline
    &  & $\langle RS \rangle$ & $P$ & $R$ & $NL$ & $H$ & $\langle RS \rangle_{k\leq10}$ & $P_{k\leq10}$ & $R_{k\leq10}$   \\
\hline {\emph{Netflix}} & HHM &0.045 & 0.062 & 0.470 & 1844 & 0.672 & 0.417 & 0.00057 & 0.018  \\
\cline{2-10}
  & BHC & 0.048 & 0.061 & 0.456 & 1777 & 0.685 & 0.443 & 0.00031 & 0.011\\
\cline{2-10}
  & HHC  & \textbf{0.041} & \textbf{0.062} & 0.453 & \textbf{1632} & \textbf{0.756} & \textbf{0.341} & \textbf{0.00347} & \textbf{0.110}\\
\hline {\emph{RYM}}
  & HHM  & 0.048 & 0.050 & 0.557 & 330 & 0.940 & 0.250 & 0.00241 & 0.092\\
\cline{2-10}
 & BHC  & 0.050 & 0.049 & 0.542 & 317 & 0.942 & 0.255 & 0.00200 & 0.081\\
\cline{2-10}
 & HHC & \textbf{0.045} & \textbf{0.051} & \textbf{0.571} & \textbf{324} & \textbf{0.947} & \textbf{0.221} & \textbf{0.00478} & \textbf{0.172}\\
\hline {\emph{MoveiLens}}
  & HHM  & 0.083 & 0.085 & 0.527 & 157 & 0.839 & 0.408 & 0.00113 & 0.044\\
\cline{2-10}
& BHC  & 0.084 & 0.084 & 0.515 & 155 & 0.839 & 0.413 & 0.00107 & 0.042\\
\cline{2-10}
  & HHC & \textbf{0.079} & \textbf{0.088} & \textbf{0.544} & \textbf{153} & \textbf{0.859} & \textbf{0.368} & \textbf{0.00202} & \textbf{0.062}\\
\hline
\end{tabular}
\begin{flushleft}The ranking score $\langle RS \rangle$, precision $P$, recall $P$, novelty $NL$, diversity $H$, object-dependent ranking score
$\langle RS \rangle_{k\leq10}$, object-dependent precision $P_{k\leq10}$ and object-dependent recall $R_{k\leq10}$ of the HHM, BHC and HHC algorithms are shown for the \emph{Netflix}, \emph{RYM} and \emph{MovieLens}, with $L=50$.
\end{flushleft}
\label{tab:2}
\end{table}

\begin{table}[!ht]
\caption{
\bf{The improvement percentage of the HHC against the HHM and BHC methods.}}
\begin{tabular}{|c|c|c|c|c|c|c|c|c|c|}
\hline
    &  & $\langle RS \rangle$ & $P$ & $R$ & $NL$ & $H$ & $\langle RS \rangle_{k\leq10}$ & $P_{k\leq10}$ & $R_{k\leq10}$   \\
\hline {\emph{Netflix}}& $\delta_{HHM}$ & 8.9\% & 0.0\% & -3.6\% &11.5\% & 12.5\% & 18.2\% & 508.8\% & 511.1\%   \\
\cline{2-10}
  & $\delta_{BHC}$ & 14.6\% & 1.6\% & -0.7\% & 8.2\% & 10.4\% & 23.0\% & 1019.4\% & 900.0\% \\
\cline{2-10}
\hline {\emph{RYM}}  & $\delta_{HHM}$ & 6.3\% & 2.0\% & 2.5\% & 1.8\% & 0.7\% & 11.6\% & 98.3\% & 87.0\% \\
\cline{2-10}
  & $\delta_{BHC}$ & 10.0\% & 4.1\% & 5.4\% & -2.2\% & 0.5\% & 13.3\% & 139.0\% & 112.3\% \\
\cline{2-10}
\hline {\emph{MovieLens}} & $\delta_{HHM}$ & 4.8\% & 3.5\% & 3.2\% & 2.5\% & 2.4\% & 9.8\% & 78.8\% & 40.9\% \\
\cline{2-10}
  & $\delta_{BHC}$ & 6.0\% & 4.8\% & 5.6\% & 1.3\% & 2.4\% & 10.9\% & 88.8\% & 47.6\% \\
\hline
\end{tabular}
\begin{flushleft}The improvement percentage of the HHC against the HHM and BHC in the ranking score $\langle RS \rangle$, precision $P$, recall $P$, novelty $NL$, diversity $H$, object-dependent ranking score
$\langle RS \rangle_{k\leq10}$, object-dependent precision $P_{k\leq10}$ and object-dependent recall $R_{k\leq10}$ is shown for the \emph{Netflix}, \emph{RYM} and \emph{MovieLens}, with $L=50$. To guide the eyes, if the HHC outperforms other methods, we show the improvement percentage as a positive value, otherwise, as a negative value.
\end{flushleft}
\label{tab:3}
\end{table}

\end{document}